\documentclass[10pt]{article}

\usepackage{amsmath}
\usepackage{amssymb}
\usepackage{graphicx}
\usepackage{cite}
\usepackage{color}
\usepackage{amsmath}
\usepackage{amssymb}
\usepackage{amsmath,amsthm,amssymb,amscd}
\usepackage{latexsym}
\usepackage{indentfirst}
\usepackage{subfigure}
\usepackage{graphicx}
\usepackage{extarrows}
\usepackage{chemarrow}
\usepackage{pifont}
\usepackage{bm}
\usepackage{multirow}

\usepackage{setspace}
\usepackage[labelfont=bf,labelsep=period,justification=raggedright]{caption}

\makeatletter
\renewcommand{\@biblabel}[1]{\quad#1.}
\makeatother

\date{}

\begin{document}

\begin{flushleft}
{\Large
\textbf{Properties of sodium-driven bacterial flagellar motor: A two-state model approach}
}
\\
Yunxin Zhang, 
\\
Shanghai Key Laboratory for Contemporary Applied Mathematics, Laboratory of Mathematics for Nonlinear Science, Centre for Computational Systems Biology, School of Mathematical Sciences, Fudan University, Shanghai 200433, China.
E-mail: xyz@fudan.edu.cn
\end{flushleft}

\section*{Abstract}
Bacterial flagellar motor (BFM) is one of the ion-driven molecular machines, which drives the rotation of flagellar filaments and enable bacteria to swim in viscous solutions. Understanding its mechanism is one challenge in biophysics. Based on previous models and inspired by the idea used in description of motor proteins, in this study one two-state model is provided. Meanwhile, according to corresponding experimental data, mathematical relationship between BFM membrane voltage and pH value of the environment, and relationship between internal and external sodium concentrations are given. Therefore, with model parameter values obtained by fitting theoretical results of torque-speed relation to recent experimental data, many biophysical properties of bacterial flagellar motor can be obtained for any pH values and any external sodium concentrations. Including the rotation speed, {\it stall torque} (i.e. the torque generated by BFM), rotation dispersion, and rotation randomness. In this study, the single-stator BFM will be firstly analyzed, and then properties of multiple-stator BFM are addressed briefly.

\section*{Author Summary}
Based on previous models for BFM and the idea used in the modeling of motor protein kinesin, one two-state model is present to discuss biophysical properties of BFM. From this model torque-speed relationship, dispersion and randomness of single-stator BFM can be obtained for any pH values and any external sodium concentrations. Meanwhile properties of BFM with multiple stators are also discussed.

\section*{Introduction}
{\it Ion-motive force} (IMF), which comprises electrical and chemical transmembrane potential, is essential for biophysical functioning of cells \cite{Hille2001, Bray2001, Howard2001, Schliwa2003}. Two primary forms of IMF are {\it proton-motive force} (PMF) and {\it sodium-motive force} (SMF). Many cellular processes, including bacterial motility, ATP synthesis and active membrane transport, are driven by SMF \cite{Shultis2006, Sowa2008, Okuno2008, Mitchell2011}. Recently, molecular machines driven by IMF have been extensively studied both experimentally and theoretically. The studies in \cite{Yasuda1998, Oster2000, Kazuhiko2000, Toyabe2011} found that, the energy efficiency of the $\rm F_1$ part of $\rm F_0F_1$-ATP synthase, which transduces energy between chemical free energy and mechanical work, is almost 100\%. In \cite{Mukherjee2011, Mukherjee2012}, coarse-grained models are employed to under the mechanism of $\rm F_1$-ATPpase and $\rm F_0$-ATPase.
Meanwhile, the catalytic power of $\rm F_1$-ATPpase is detailed analyzed in \cite{Adachi2007}, the structure and organization of the yeast $\rm F_0F_1$-ATP synthase are studied in \cite{Davies2012}, and the Na$^+$, K$^+$-ATPase in liposomal membrane are discussed in \cite{Bouvrais2012}. Finally, more theoretical studies can also be found in \cite{Xing2004, Diez2004, Berry2005, Sielaff2008, Junge2009, Hayashi2010, Toyabe2010, Rieko2010, Little2011, Goldstein2011, Kohori2011, Sugawa2011, Iino2012, Tanigawara2012, Okazaki2013} for $\rm F_0$-ATPase or $\rm F_1$-ATPase, and in \cite{Choua1999, Berezhkovskii2002, Berezhkovskii2003, Berezhkovskii2005, Berezhkovskii20051, Bezrukov2007, Kolomeisky20071, Kolomeisky2008, Zilman2009}
for transmembrane motion of ions.

One of the most important molecular machines driven by IMF is bacterial flagellar motor (BFM), which couples ion flow ($\rm H^+$ or $\rm Na^+$) to the rotation of extracellular helical flagellar filaments at hundreds of revolutions per second (Hz), and then propels many species of swimming bacteria \cite{Stock2012}. Using the transmembrane electrochemical H$^+$(or Na$^+$) gradient to power rotation of the flagellar motor, free-swimming bacteria can propel their cell body at a speed of 15-100 mm/s, or up to 100 cell body lengths/s \cite{Berg2003, Sowa2008}. The BFM has one rotor and multiple stators in a circular ring-like structure $\approx45$ nm in diameter \cite{Thomas2006}. The stators are attached to the rigid peptidoglycan cell wall and the spinning of the rotor drives the flagellar filament through a short hook. In {\it E. coli}, the rotor is composed of a ring of $\sim$26 FliG proteins and each stator has 4 copies of proteins MotA and 2 copies of proteins MotB, forming 2 proton-conducting transmembrane channels \cite{Meacci2009, Bai2012}. The stator can deliver torque to the rotor by converting the free energy of ion flow  across the cytoplasmic membrane.

To understand the biophysical properties and torque-generating mechanism of BFM, many experiments have been done to measure revolution speed under varying torque \cite{Ryu2000, Chen2000, Gabel2003, Sowa2003, Asai2003, Bouvrais2008, Inoue2008, Nakamura2009, Yuan2010}. Meanwhile, various biophysical and biochemical models have also been constructed to try to understand the basic principle of BFM \cite{Ryu2000, Schmitt2003, Xing2006, Mora2009, Meacci2009, Bai2009, Bai2012}. However, so far most of the previous studies mainly focused on the torque-speed relationship, except in \cite{Mora2009, Nakamura2009} where effect of pH value on torque-speed relation is discussed, and in \cite{Bai2009} where effects of pH value and membrane potential are analyzed theoretically, and in \cite{Chen2000, Gabel2003, Xing2006, Meacci2009} where the effect of temperature (or noise) on the speed of BFM is addressed, and in \cite{Sowa2003, Asai2003} where the effect of Na$^+$ concentration is studied experimentally.

\begin{figure}[!ht]
\begin{center}
\includegraphics[width=6in]{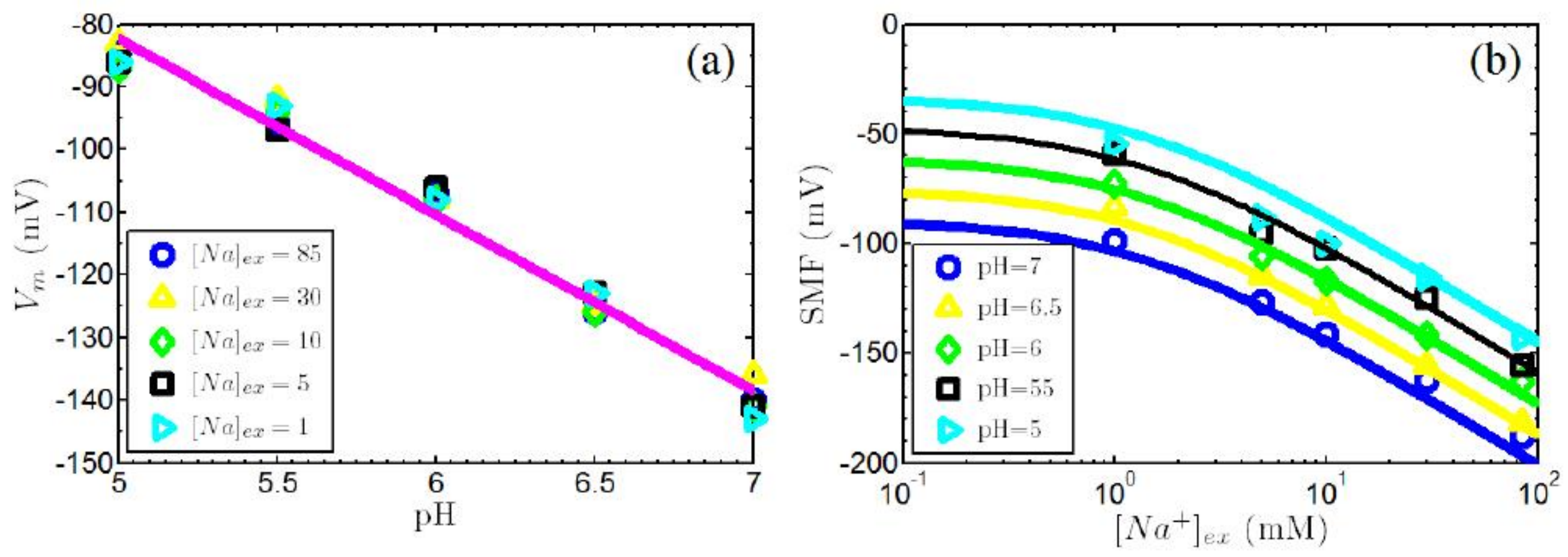}
\end{center}
\caption{{\bf (a)} The relationship between pH value and membrane voltage $V_m$, which is used to get electrical energy $U=qV_m$, can be well approximated by $V_m=-28.04{\rm pH}+57.88$, see Eq. (\ref{eq1}). {\bf (b)} The {\it sodium-motive force} (SMF) can be approximately obtained by ${\rm SMF}=V_m+k_BT/q\ln[9.013/(1.318+{\rm [Na]}_{ex})]$, which means the internal sodium concentration ${\rm [Na]}_{in}$ can be obtained by the external sodium concentration ${\rm [Na]}_{ex}$, ${\rm [Na]}_{in}=9.013{\rm [Na]}_{ex}/(1.318+{\rm [Na]}_{ex})$, see Eq. (\ref{eq7}). In this figure, all the data are from \cite{Lo2013}.
}
\label{FIGpHNaInterpolation}
\end{figure}

In recent study of Lo {\it et al} \cite{Lo2013}, one three-state model was presented to describe the torque-speed relationship of one-stator BFM under various pH values and various external sodium concentrations. Their model can fit experimental data well, in which one of the basic parameters is the electrical energy $U=qV_m$, with $V_m$ the membrane voltage. Experimental data shows that membrane voltage $V_m$ depends on surrounding solution (see \cite{Lo2013} and references there in). This study found that, the value of $V_m$ depends only on the pH value of the solution, and is independent of the external sodium concentration [Na]$_{ex}$ [see Fig. \ref{FIGpHNaInterpolation}{\bf (a)}]. Moveover, $V_m$ changes almost linearly with pH value,
\begin{equation}\label{eq1}
V_m=-28.04{\rm pH}+57.88,
\end{equation}
with voltage in millivolt unit (mV).
On the other hand, besides pH value and external sodium concentration [Na]$_{ex}$, the value of internal sodium concentration [Na]$_{in}$ is also needed to obtain the {\it sodium-motive force}, since it comprises electrical transmembrane potential $V_m$ and chemical transmembrane potential $\Delta\mu/q$ and is defined as
\begin{equation}\label{eq2}
{\rm SMF}=V_m+\frac{\Delta\mu}{q}=V_m+\frac{k_BT}{q}\ln\left(\frac{{\rm [Na]}_{in}}{{\rm [Na]}_{ex}}\right),
\end{equation}
where $q=1.602\times10^{-19}$ coulomb (C) is the charge of sodium ion, $k_B$ is the Boltzmann constant, and $T$ is the absolute temperature. Further analysis about the experimental data shows that the internal sodium concentration [Na]$_{in}$ can be well approximated by [see Fig. \ref{FIGpHNaInterpolation}{\bf (b)}]
\begin{equation}\label{eq3}
{\rm [Na]}_{in}=9.013{\rm [Na]}_{ex}/(1.318+{\rm [Na]}_{ex}).
\end{equation}
In fact, this relation is similar as the one used to approximate ADP concentration by ATP concentration in the study of motor protein kinesin \cite{Fisher2001}.

Different with the three-state model used in \cite{Lo2013}, in this study one simple two-state model will be presented to describe the rotation of one-stator BFM. With relationships (\ref{eq1}) and (\ref{eq3}), properties for one-stator BFM under any pH values and external sodium concentrations will be obtained, such as the torque-speed relationship, the {\it stall torque} $\Gamma_s$ (i.e. the torque under which the rotation speed $V$ of BFM vanishes), dispersion $D$ and randomness $r$. The results show that, both rotation speed $V$ and {\it stall torque} $\Gamma_s$ increase with pH value and sodium concentration [Na]$_{ex}$. Under high pH value and sodium concentration [Na]$_{ex}$, the rotation of one-stator BFM can be regarded as one Possion process, i.e. process with only one internal state. Meanwhile, in this study properties of BFM with multiple stators will also be discussed.

\section*{Results}
\begin{figure}[!ht]
\begin{center}
\includegraphics[width=6in]{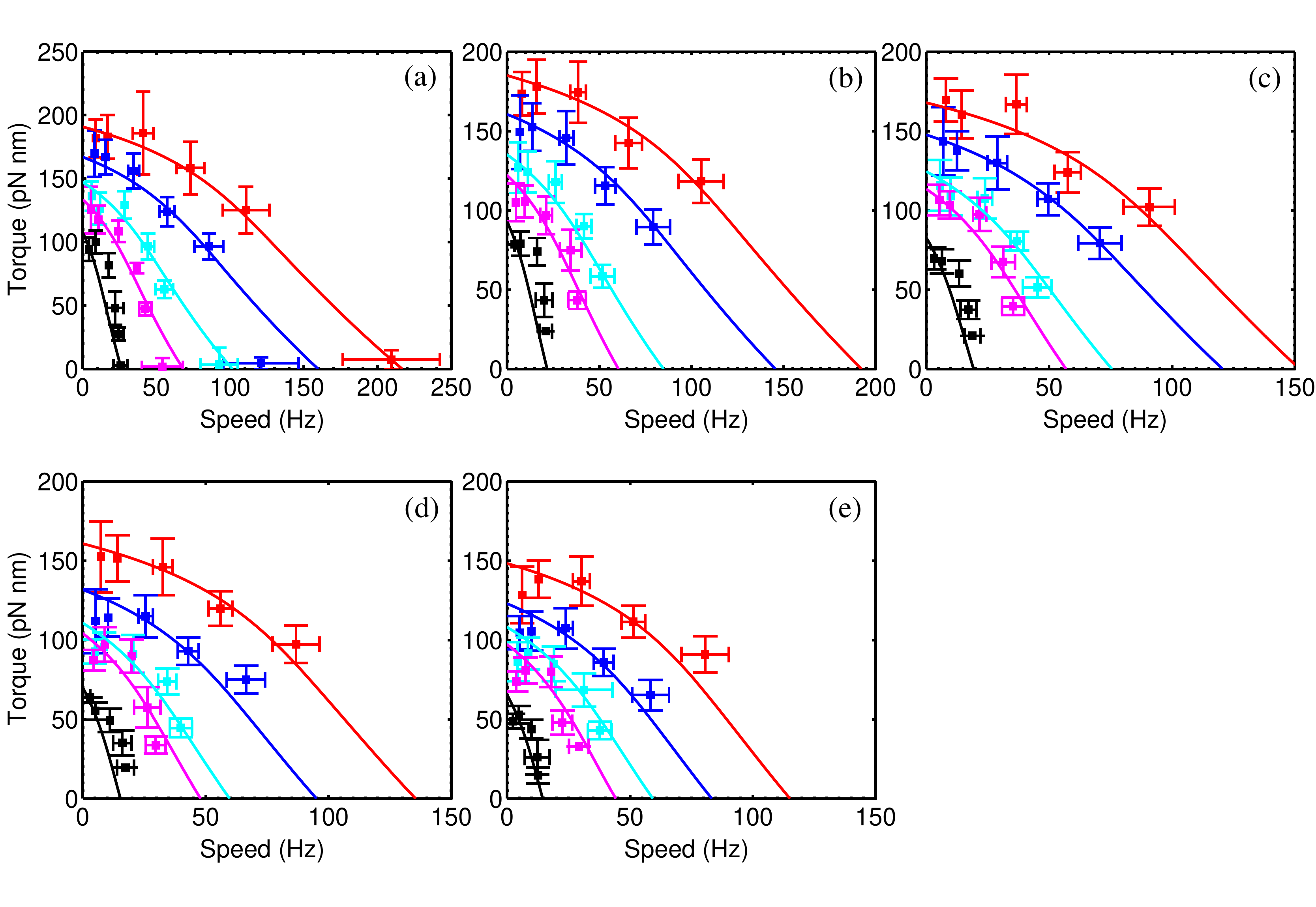}
\end{center}
\caption{Torque-speed relationships of one-stator BFM for {\bf (a)} pH=7, {\bf (b)} pH=6.5, {\bf (c)} pH=6, {\bf (d)} pH=5.5, and {\bf (e)} pH=5. From right to left, the lines are for external sodium concentration [Na]$_{ex}$=85 mM, 30 mM, 10mM, 5 mM, and 1 mM respectively. The experimental data are from \cite{Lo2013}, and the lines are obtained from the two-state model [see Eq. (\ref{eq6})] with model parameter values listed in Table \ref{Tab1}.
}
\label{FIGFitResults}
\end{figure}

In previous studies, four-state model is usually used to describe the rotation of BFM \cite{Sowa2003, Inoue2008, Nakamura2009}. However, such model has 25 free model parameters which makes it difficult to fit corresponding data, and special Monte Carlo search strategy should be given to overcome this difficulty \cite{Lo2013}. For simplicity, and inspired by the corresponding idea used in the modeling of motor proteins \cite{Fisher2001, Kolomeisky2007}, one two-state model is presented in this study. Compared with the four-state model, this two-stator model has only 10 free parameters, which is even less than that of the reduced three-state model as used in \cite{Lo2013}. Moreover, by this two-state model, it is much convenient to do further theoretical analysis to know more biophysical properties about the rotation of BFM. For example, the dispersion $D$ and randomness $r=2D/V\varphi$ can be easily obtained. At the same time, by the relationship between membrane voltage $V_m$ and pH value as given in Eq. (\ref{eq1}), and the relationship between internal sodium concentration [Na]$_{in}$ and external sodium concentration [Na]$_{ex}$ as given in Eq. (\ref{eq3}), properties of BFM rotation can be obtained for any pH values and any external sodium concentrations (in the recent paper of Lo {\it et al} \cite{Lo2013}, only torque-speed relationships for pH=5, 5.5, 6, 6.5, 7, and external sodium concentration [Na]$_{ex}$=1 mM, 5 mM, 10 mM, 30 mM, 85 mM are provided, see also Fig. \ref{FIGFitResults}). The fitting results given in Fig. \ref{FIGFitResults} show that, with appropriate parameter values (see TABLE \ref{Tab1}), this two-state model can explain experimental data reasonably.

\begin{figure}[!ht]
\begin{center}
\includegraphics[width=6in]{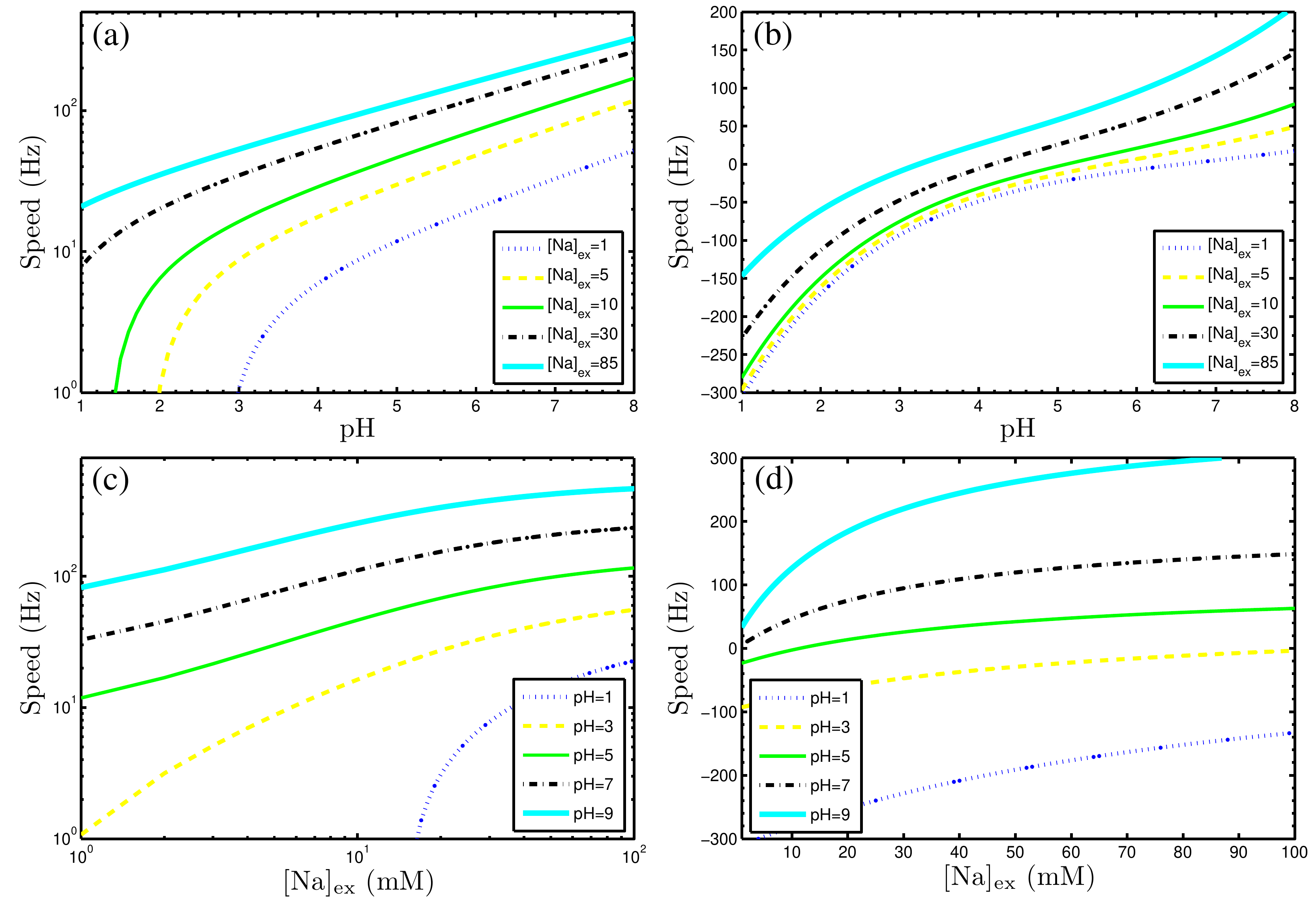}
\end{center}
\caption{The rotation speed $V$ of one-stator BFM increases with pH value and external sodium concentration [Na]$_{ex}$. For high pH value, speed $V$ increases almost experimentally with pH. In {\bf (a, c)}, the torque $\Gamma=0$, and in {\bf (b, d)}, the torque $\Gamma=100$ pN$\cdot$nm. 
  For high values of $\Gamma$ and low values of pH and sodium concentration [Na]$_{ex}$, the stator will rotate reversely.
}
\label{FIGProperty}
\end{figure}

As mentioned previously, using formulations given in Eqs. (\ref{eq1}) and (\ref{eq3}), properties of BFM rotation can be obtained for any pH values and any external sodium concentrations [Na]$_{ex}$. The plots in Fig. \ref{FIGProperty} show that, under any external torques BFM rotation speed $V$  increases monotonically with pH value and external sodium concentration [Na]$_{ex}$. For high concentration [Na]$_{ex}$ and pH values, $V$ increases almost experimentally with pH value.
Further calculations indicate that, with the increase of sodium concentration [Na]$_{ex}$, the rotation speed $V$ will tend to one limit value. Since for high external concentration [Na]$_{ex}$, the internal concentration [Na]$_{in}$, and consequently the transition rate $k_4$, is almost invariable [see Eq. (\ref{eq3})]. So the rotation speed as given in Eq. (\ref{eq6}) satisfies $V\to k_3\varphi$.

\begin{figure}[!ht]
\begin{center}
\includegraphics[width=6in]{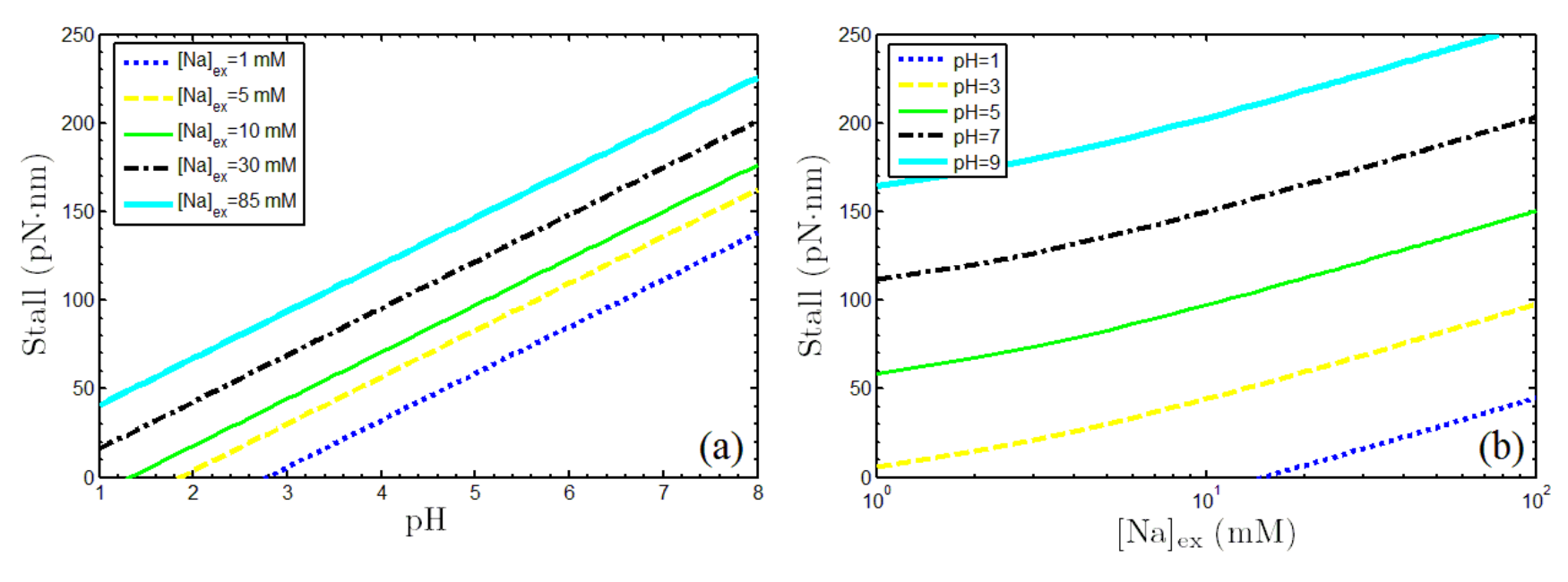}
\end{center}
\caption{The {\it stall torque} $\Gamma_s$, i.e. the torque under which the rotation speed of BFM vanishes, increases with pH value (linearly) and external sodium concentration [Na]$_{ex}$, see the formulation in Eq. (\ref{eq7}) [{\bf Note}, the figures are obtained directly by $V(\Gamma_s)=0$, see the corresponding formulation of speed $V$ in Eq. (\ref{eq6})].
}
\label{FIGStall}
\end{figure}

The plots in Fig. \ref{FIGStall} show that, the {\it stall torque} $\Gamma_s$, i.e. the torque which satisfies $V(\Gamma_s)=0$, increases linearly with pH value and almost linearly with the logarithm of external sodium concentration $\ln$[Na]$_{ex}$, see also Eq. (\ref{eq7}). So concentrations of proton and sodium ion not only have effect on rotation speed $V$, but also can change the maximum output work $W_{\max}=\Gamma_s\varphi$ \cite{Fisher19991, Toyabea2012}. Further calculations show that other biophysical quantities, such as the maximum output power $P_{\max}=\max_{\Gamma}\Gamma V(\Gamma)$, and efficiency at maximum output power $\eta_{\max}=\Gamma_{\max}/\Gamma_s$, depend also on pH value and sodium concentration [Na]$_{ex}$ \cite{Zhang2013}. Where $\Gamma_{\max}$ is the external torque under which the output power of BFM reaches its maximum $P_{\max}$.

\begin{figure}[!ht]
\begin{center}
\includegraphics[width=6in]{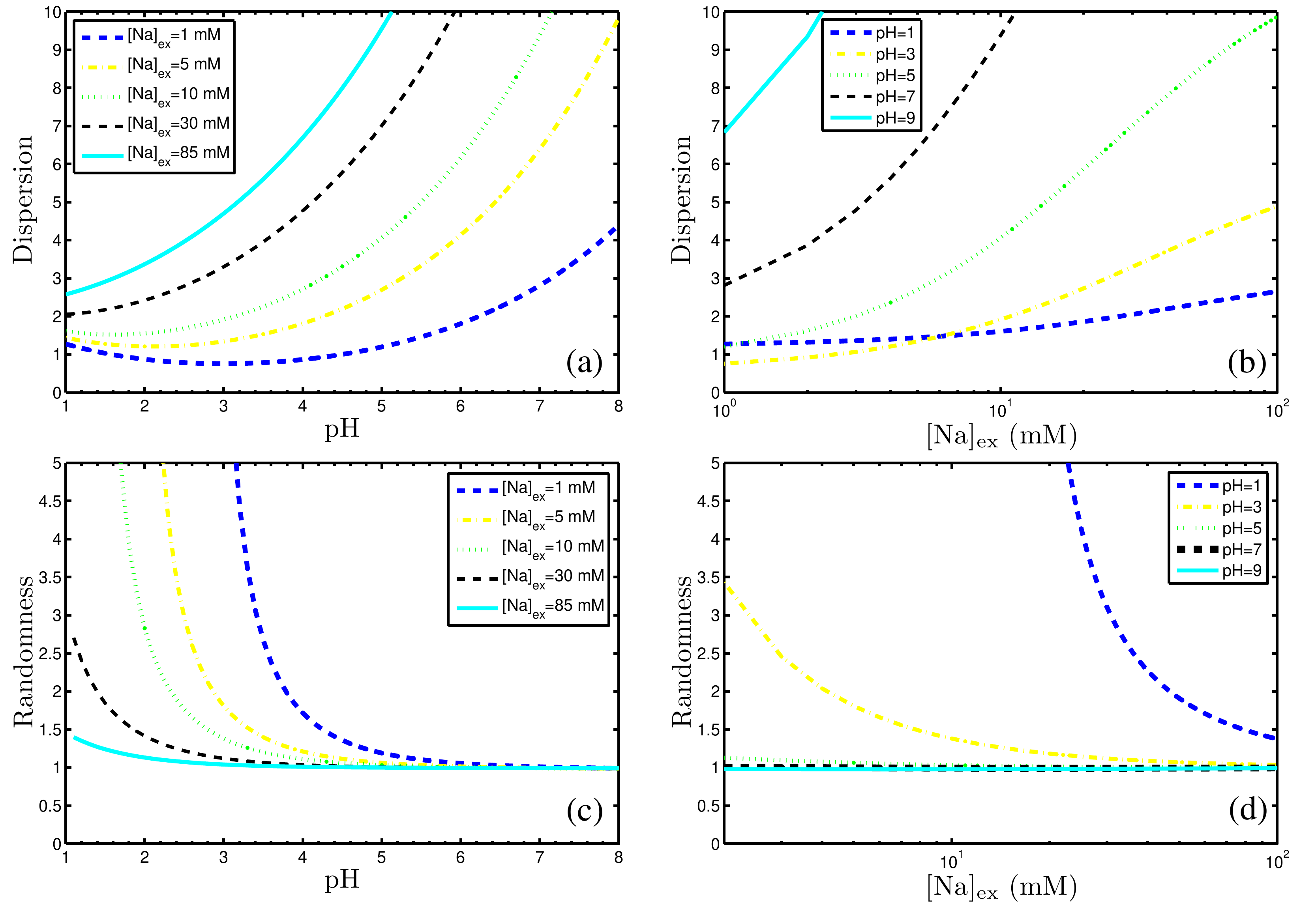}
\end{center}
\caption{The dispersion $D$ and randomness $r$ of BFM rotation. {\bf (a)} For high sodium concentration, dispersion $D$ increases with pH value ({\bf note}, for low sodium concentration and low pH values, the rotation speed $V$ is negative, see Fig. \ref{FIGProperty}); {\bf (b)} Dispersion $D$ increases with sodium concentration [Na]$_{ex}$. With the increase of pH value and sodium concentration, randomness $r=2D/V\varphi$ decreases monotonically to 1 {\bf (c, d)}. Which means that, for high pH value and [Na]$_{ex}$, the rotation of BFM is just one Possion process. In all figures, the external torque is zero.
}
\label{FIGDandRandomF0}
\end{figure}

Besides rotation speed $V$ and {\it stall torque} $\Gamma_s$, the dispersion $D$ of BFM rotation also increases with pH value and external sodium concentration [Na]$_{ex}$, see Fig. \ref{FIGDandRandomF0}({\bf a, b}) . Here dispersion $D$, also called {\it effective diffusion constant}, is defined by $D=1/2\lim_{t\to\infty}d[\langle\theta^2(t)\rangle-\langle\theta(t)\rangle^2]/dt$ \cite{Derrida1983, Fisher2001, Zhang20092}.
However, with the increase of pH value and external sodium concentration [Na]$_{ex}$, the randomness $r=2D/V\varphi$ decreases to 1 monotonically. Which means that, under high pH value and high sodium concentration [Na]$_{ex}$, the BFM rotation is just one Possion process \cite{Visscher1999}. As discussed before, in such cases, the forward transition ${\bf 1}\xrightarrow[]{k_1}{\bf 2}$ is fast, and the rotation is mainly limited by the forward transition ${\bf 2}\xrightarrow[]{k_3}{\bf 1}$, see Fig. \ref{FIGschematic}({\bf b}).

\begin{figure}[!ht]
\begin{center}
\includegraphics[width=6in]{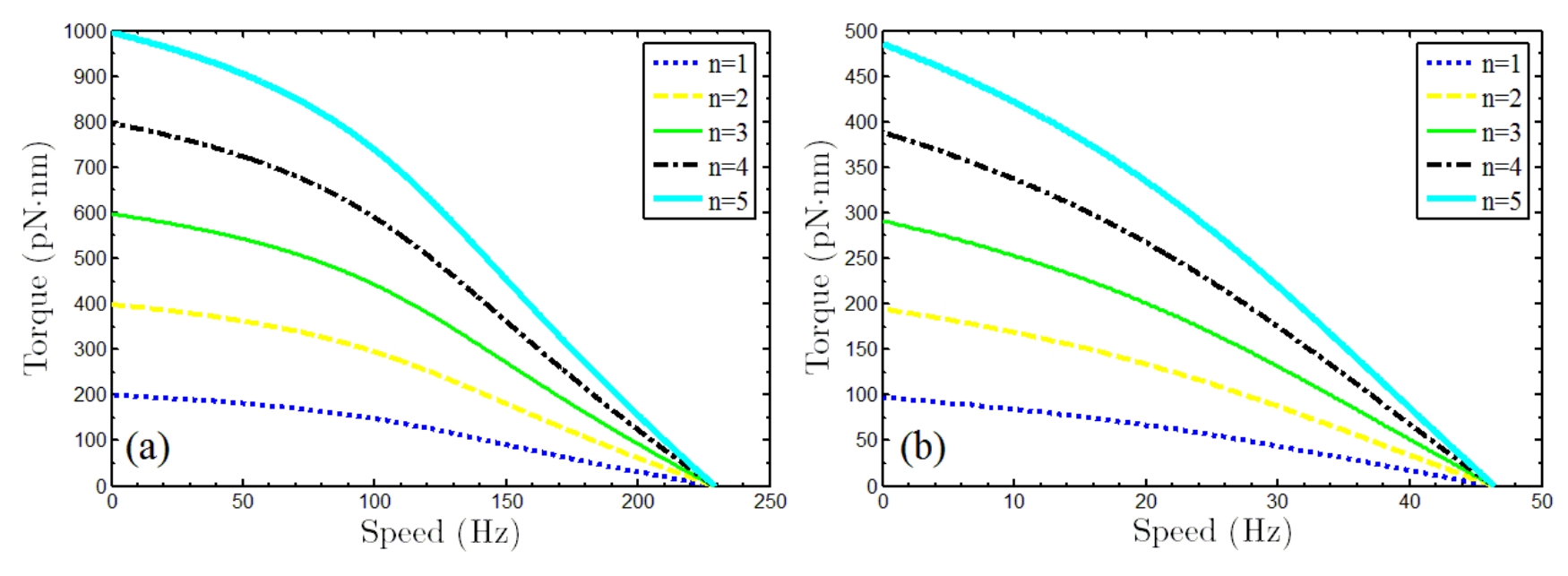}
\end{center}
\caption{Torque-speed relationships if the external torque $\Gamma$ is shared equally by $n$ stators for $n=1, 2, 3, 4, 5$ respectively. {\bf (a)} pH=7 and [Na]$_{ex}=85$ mM, {\bf (b)} pH=5 and [Na]$_{ex}=10$ mM. The figures are similar as the ones obtained in \cite{Ryu2000, Meacci2009}.
}
\label{FIGMultipleMotor}
\end{figure}

Since one BFM may include multiple stators, the external torque is then shared by all the working stators. So the {\it stall torque} of BFM with $n$ stators should be $n$ times of which of the single stator BFM, while the maximum rotation speeds (i.e. speeds without external torque) are the same \cite{Ryu2000, Meacci2009, Lo2013}, see Fig. \ref{FIGMultipleMotor}. However, for a BFM with $M$ stators, the $M$ stators may not work all the time. In other words, some stators may be dissociated from the rotor of BFM. So they are in idle state, or their rotation have no effective contribution to the BFM rotation. Therefore, for multiple stator BFM, the torque-speed relationship may be different from the ones plotted in Fig. \ref{FIGMultipleMotor}. Actually, the actual rotation speed is slightly lower than that plotted in Fig. \ref{FIGMultipleMotor} (figures not shown here). In the following, the properties of BFM with multiple stators will be discussed.

\begin{figure}[!ht]
\begin{center}
\includegraphics[width=6in]{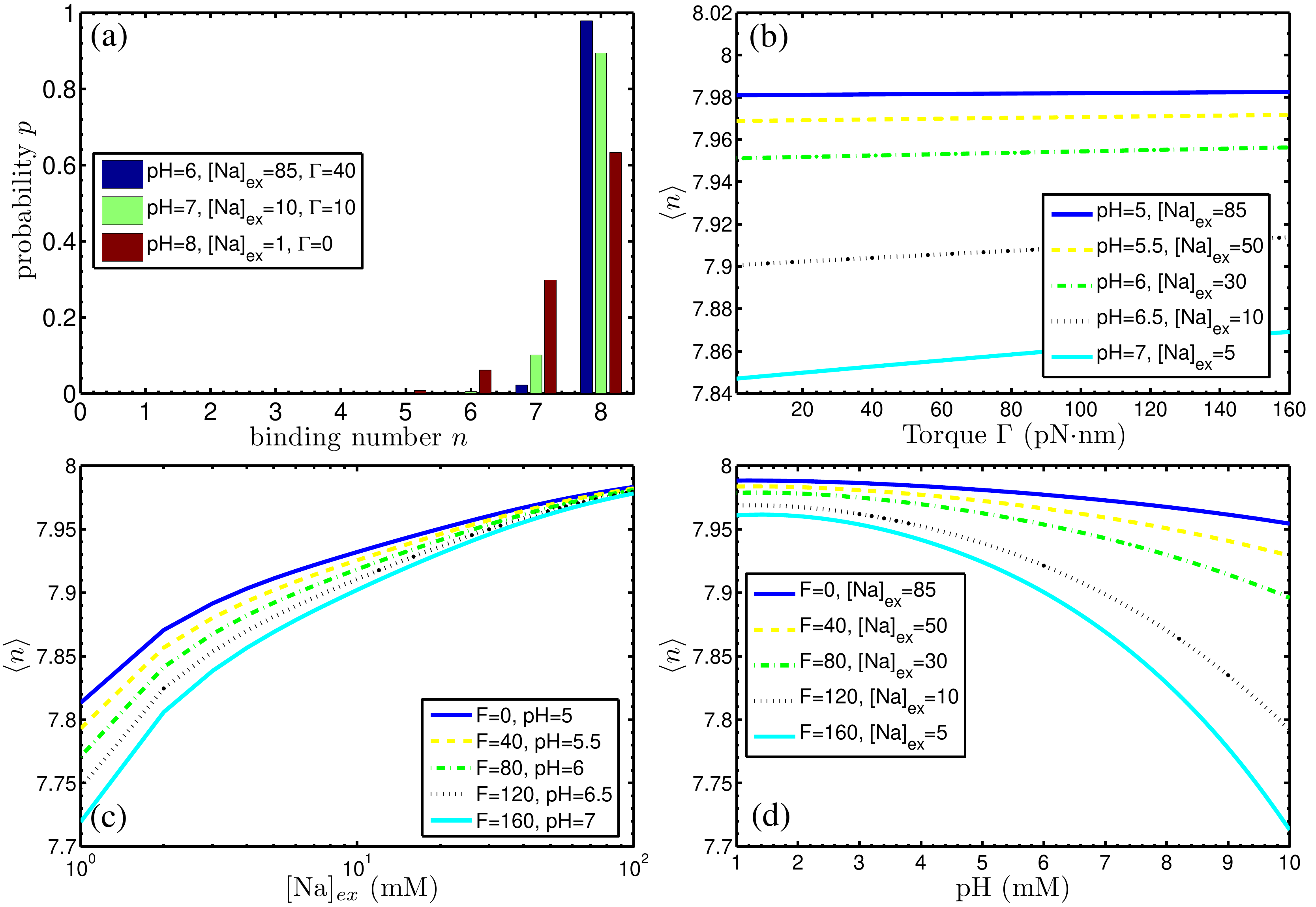}
\end{center}
\caption{{\bf (a)} Probability of stator number binding to BFM rotor. In most cases, almost all stators are bound to BFM rotor, i.e. the duty ratio of stator is high \cite{Ryu2000}. {\bf (b)} With the increase of external torque $\Gamma$, the mean binding number $\langle n\rangle$ of stator will be increased. Meanwhile, the mean binding number $\langle n\rangle$ increases with sodium concentration [Na]$_{ex}$ {\bf (c)}, but decreases with pH value {\bf (d)}. In all the figures, the total number of stator is assumed to be $M=8$.
}
\label{FIGProbabilityOfStatorNumber}
\end{figure}

Three examples of probability $p_n$ that there are $n$ stators in working to rotate the BFM are given in Fig. \ref{FIGProbabilityOfStatorNumber}({\bf a}) for a BFM with total $M=8$ stators. Although most of stators are engaged in the rotation of BFM, some may in unbinding state, especially when the external torque is low. The plots in Fig. \ref{FIGProbabilityOfStatorNumber}({\bf b}) show that the mean binding number $\langle n\rangle$ of stator increases slightly with the external torque $\Gamma$, but is always less than the total stator number $M$. This indicates that, the BFM can adjust itself with the external torque, if the torque is low some stators may in idle state to save energy. Meanwhile, the mean binding number $\langle n\rangle$ increases with sodium concentration [Na]$_{ex}$ [see Fig. \ref{FIGProbabilityOfStatorNumber}({\bf c})]. Since with high sodium concentration [Na]$_{ex}$ the transition rates $k_1, k_4$ are high [see Fig. \ref{FIGschematic}({\bf b}) and discussion below Eq. (\ref{eq5})], consequently stators are more likely to transfer to the binding state {\bf 2}. Finally, the mean binding number $\langle n\rangle$ decreases with the pH value, see Fig. \ref{FIGProbabilityOfStatorNumber}({\bf d}).

\begin{figure}[!ht]
\begin{center}
\includegraphics[width=6in]{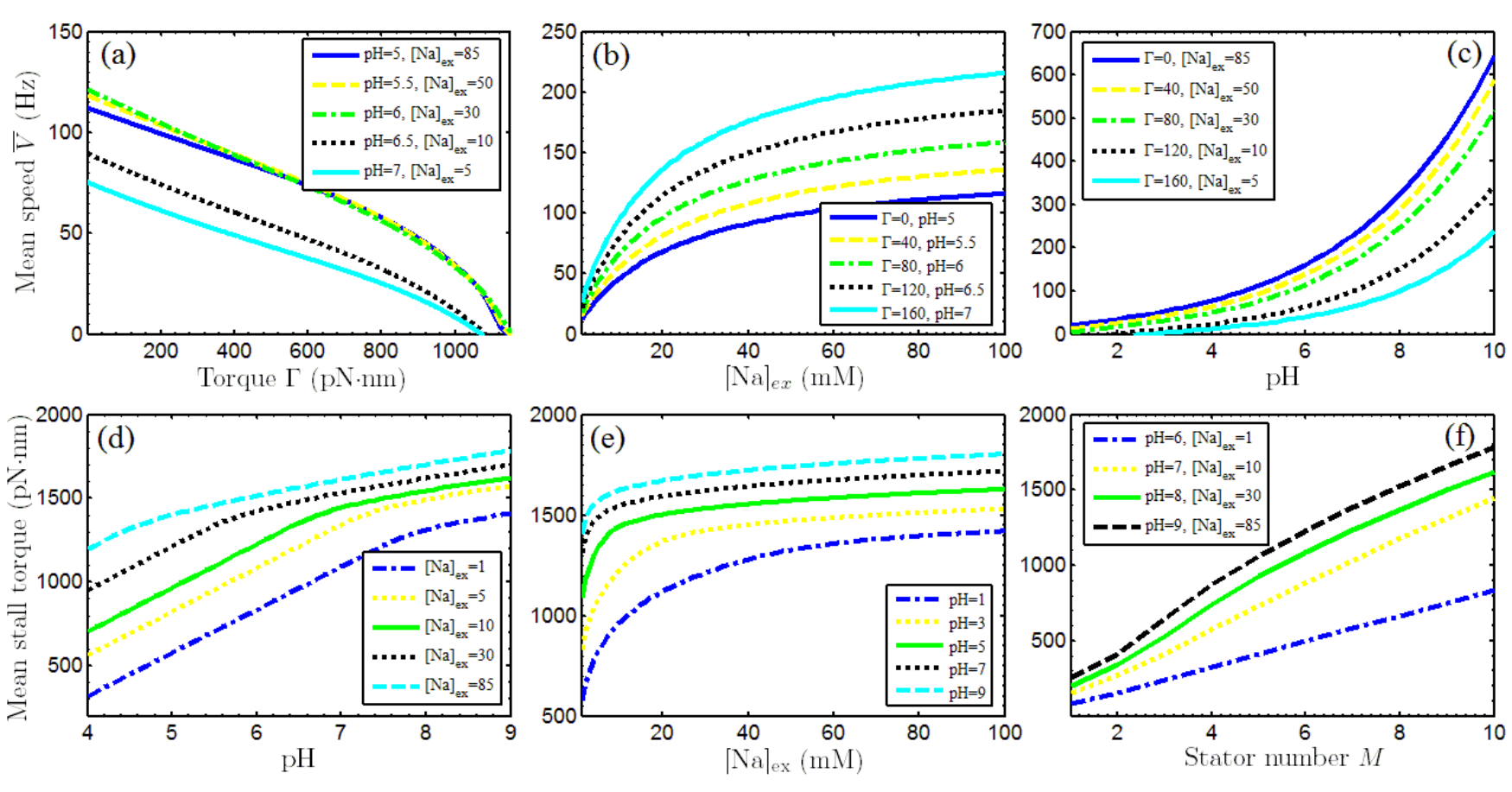}
\end{center}
\caption{The {\it mean rotation speed} $\overline{V}$ and {\it mean stall torque} $\overline{\Gamma}_s$ of BFM with total stator number $M=8$. Here $\overline{V}$ is obtained by $\overline{V}=\sum_{n=1}^Mp_nV_n$, in which $p_n$ is the probability that BFM rotor is bound by $n$ stators, and $V_n$ is the rotation speed of BFM if the external torque is shared by $n$ stators. The {\it mean stall torque} $\overline{\Gamma}_s$ is obtained by $\overline{V}(\overline{\Gamma}_s)=0$. Similar as the one single stator cases, the {\it mean rotation speed} $\overline{V}$ decreases with external torque $\Gamma$ {\bf (a)}, and increases with external sodium concentration [Na]$_{ex}$ and pH value {\bf (b, c)}. Meanwhile, the {\it mean stall torque} $\overline{\Gamma}_s$ increases with [Na]$_{ex}$, pH value and total stator number $M$ {\bf (d, e, f)}.
}
\label{FIGMeanVandStall}
\end{figure}

As mentioned before, usually the mean rotation speed $\overline{V}=\sum_{n=1}^Mp_nV_n$ of a BFM with $M$ stators is lower than its maximum speed $V_M$ [the BFM speed provided all the $M$ stators are bound to the BFM rotor (see Fig. \ref{FIGMultipleMotor} for $V_M$)]. But the plots in Fig. \ref{FIGMeanVandStall}({\bf a,b,c}) show that, similar as $V_M$, mean rotation speed $\overline{V}$ also decreases with external torque $\Gamma$, and increases with sodium concentration [Na]$_{ex}$ and pH value [also experimentally for large [Na]$_{ex}$ and pH value, Fig. \ref{FIGProperty}({\bf a})]. This is because $\overline{V}$ is merely the linear combination of $V_n$ for $n=1,2,\cdots,M$. As expected, the mean {\it stall torque} $\overline{\Gamma}_s$ increases with sodium concentration [Na]$_{ex}$ and pH value, and increased almost linearly with the total stator number $M$.

\section*{Discussion}
In this study, the sodium driven bacterial flagellar motor (BFM) is theoretically analyzed. Based on the previous models used in \cite{Sowa2003, Inoue2008, Nakamura2009, Lo2013} and the idea used in \cite{Fisher2001, Kolomeisky2007} for modeling description of motor protein kinesin, one simple two-state model is presented. Moreover, according to the experimental data collected in \cite{Lo2013}, relationship between membrane voltage and pH value, and relationship between internal and external sodium concentrations are explicitly given. So, by the two-state model, rotation speed of BFM can be obtained for any pH values and any external sodium concentrations. Meanwhile, in this study other biophysical properties of BFM are also studied, including {\it stall torque} (the torque under which the BFM rotation speed vanishes), dispersion and randomness. In reality, one BFM may have multiple stators. In this study, the properties of multiple-stator BFM are also addressed. Roughly speaking, the {\it stall torque} of a $M$-stator BFM is about $M$ times of which of a single-stator BFM, and the external torque of BFM are usually shared by all the $M$ stators. But under low torque, low sodium concentration and high pH value, some stators may be disassociated from BFM rotor and in idle state. Using this two-state model, other detailed biophysical properties can also be obtained, such as the maximum power and energy efficiency \cite{Zhang20091, Toyabea2012, Zhang2013}.

\section*{Materials and Methods}
\subsection*{Two-state kinetic model for one-stator BFM}
\begin{figure}[!ht]
\begin{center}
\includegraphics[width=6in]{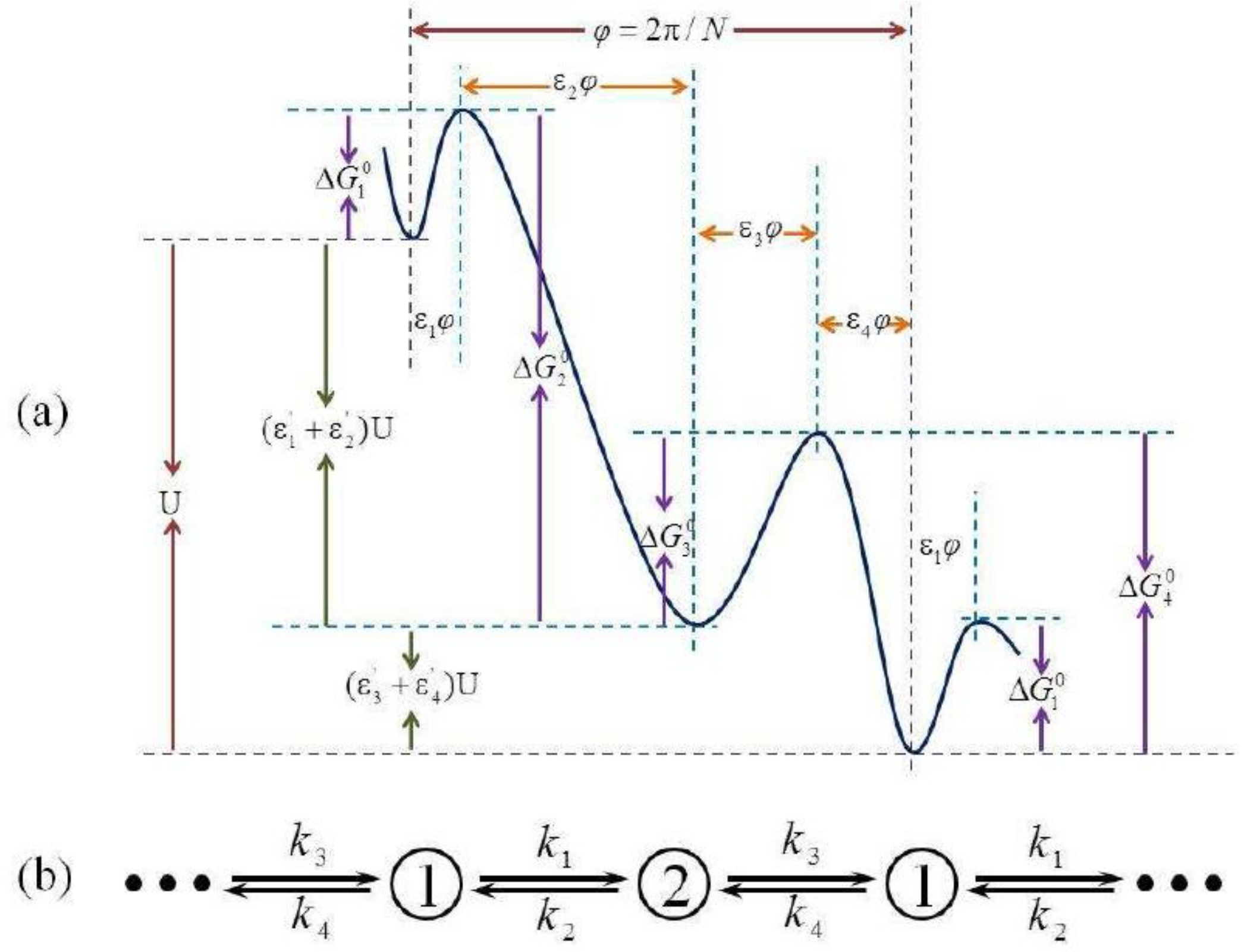}
\end{center}
\caption{Schematic depiction of the two-state model for stator rotation. {\bf (a)} The stator is assumed to move in one tilted periodic potential $G(\theta)$ with period $\varphi=2\pi/N$. Where $N=37$ is the number of ions which cross the membrane in each revolution of bacterial flagellar motor \cite{Lo2013}. Under no external torque, the energy difference in one period is equal to the electrical energy $U$ released by one ion crossing membrane. The {\it distribution factor}s $\epsilon_i$ and $\epsilon_i'$ satisfy the normalization condition $\sum_{i=1}^4\epsilon_i=\sum_{i=1}^4\epsilon_i'=1$, and for simplicity they are assumed to be independent of external torque $\Gamma$ and the electrical energy $U$ \cite{Zhang2013}. With the increase of external torque, the forward rotation barriers (in this figure $\Delta G_1^0, \Delta G_3^0$ are the energy barriers when external torque vanishes) will be increased, while the backward rotation barriers  (similarly, $\Delta G_2^0, \Delta G_4^0$ are the backward energy barriers  when external torque vanishes) will be decreased. Consequently, the degree of the potential tilt will be decreased. The two states of the stator are corresponding to the two minima of the potential. $\epsilon_i\varphi$ is the mechanical angle between stable state of the stator and its rotation energy barrier. The rotation speed of stator increases with electrical energy $U$  but decreases with external torque $\Gamma$. {\bf (b)} Mathematically, the stator rotation in the periodic tilted potential depicted in {\bf (a)} can be described by one Markov process with two biochemical state, denoted by state {\bf 1} and state {\bf 2} respectively. In which one forward biochemical cycle ${\bf 1}\xrightarrow[]{k_1}{\bf 2}\xrightarrow[]{k_3}{\bf 1}$ is coupled with one forward mechanical rotation of angle $\varphi$. The transition rates $k_i$ depend on electrical energy $U$, external torque $\Gamma$, ion concentration and pH value, which are given by formulations in (\ref{eq5}).
}
\label{FIGschematic}
\end{figure}

The two-state model used in this study, which is schematically depicted in Fig. \ref{FIGschematic}({\bf a}), can be regarded as one combination of the three-state model presented in \cite{Lo2013} for BFM and the two-state model given in \cite{Fisher2001} for motor protein kinesin. In brief, the rotation of stator can be regarded as motion of a particle in one tilted periodic potential $G(\theta)$ with period $\varphi=2\pi/N$, i.e. periodic rotation with step-size $\varphi$. Where $N$ is the number of ions passing through  stator per revolution, and as in \cite{Lo2013}, $N=37$ is used in this study. The same as in the modeling for motor protein kinesin \cite{Fisher2001}, in each period potential $G(\theta)$ has two minima, corresponding to two transition states of stator when one ion passes through it. Or in other words, the stator has two states in each rotation step (with step-size $\varphi$), denoted by {\bf 1} and {\bf 2} respectively. Mathematically, the periodic rotation of stator can be simply described by the following Markov process (see also Fig. \ref{FIGschematic}({\bf b}))
$$
\cdots\Large{\autorightleftharpoons{$k_3$}{$k_4$}} \Large{\textcircled{{\small\bf 1}}}\autorightleftharpoons{$k_1$}{$k_2$}\Large{\textcircled{{\small\bf 2}}}\autorightleftharpoons{$k_3$}{$k_4$}\Large{\textcircled{{\small\bf 1}}}\autorightleftharpoons{$k_1$}{$k_2$}\cdots.
$$
During the forward rotation from state {\bf 1} to state {\bf 2}, stator should overcome energy barrier $\Delta G_1$, so the transition rate $k_1\propto e^{-\Delta G_1/k_BT}$. Similarly, $k_i\propto e^{-\Delta G_i/k_BT}$ for $i=2, 3, 4$. Where energy barrier $\Delta G_i$ depends on external torque $\Gamma$. Similar to the method as demonstrated in models of molecular motor and microtubule \cite{Fisher2001, Zhang20093, Zhang20112}, this torque dependence can be approximated as follows
\begin{equation}\label{eq4}
\Delta G_1=\Delta G_1^0+\epsilon_1\Gamma\varphi,\quad \Delta G_2=\Delta G_2^0-\epsilon_2\Gamma\varphi,\quad
\Delta G_3=\Delta G_3^0+\epsilon_3\Gamma\varphi,\quad \Delta G_4=\Delta G_4^0-\epsilon_4\Gamma\varphi.
\end{equation}
Where $\Delta G_i^0$ (for $i=1,2,3,4$) are intrinsic energy barriers, i.e. energy barriers when external torque $\Gamma$ vanishes. Parameters $\epsilon_i$ are {\it load distribution factor}s which satisfy normalization condition $\sum_{i=1}^4\epsilon_i=1$. The electrical energy $U$ released by one ion when passing through stator can be decomposed into two parts, $U=U_1+U_2:=(\Delta G_2^0-\Delta G_1^0)+(\Delta G_4^0-\Delta G_3^0)$, with $U_1$ the energy released during forward transition ${\bf 1}\xrightarrow[]{k_1}{\bf 2}$, and $U_2$ the energy released during forward transition ${\bf 2}\xrightarrow[]{k_3}{\bf 1}$. Based on the above analysis, in the two-state kinetic model, transition rates $k_i$ (see Fig. \ref{FIGschematic}({\bf b})) can be obtained by the following formulations,
\begin{equation}\label{eq5}
k_1=k_1^0e^{(\epsilon_1'U-\epsilon_1\Gamma\varphi)/k_BT},\quad
k_2=k_2^0e^{(-\epsilon_2'U+\epsilon_2\Gamma\varphi)/k_BT},\quad
k_3=k_3^0e^{(\epsilon_3'U-\epsilon_3\Gamma\varphi)/k_BT},\quad
k_4=k_4^0e^{(-\epsilon_4'U+\epsilon_4\Gamma\varphi)/k_BT}.
\end{equation}
Where $\epsilon'_i$ are also {\it distribution factor}s which satisfy $(\epsilon'_1+\epsilon'_2)=U_1/U$, and $(\epsilon'_3+\epsilon'_4)=U_2/U$. Energies $\epsilon_1'U, \epsilon_3'U$ are the parts to accelerate forward transitions ${\bf 1}\xrightarrow[]{k_1}{\bf 2}$ and ${\bf 2}\xrightarrow[]{k_3}{\bf 1}$ respectively. While $\epsilon_2'U, \epsilon_4'U$ are energies to reduce backward transitions ${\bf 1}\xleftarrow[k_2]{}{\bf 2}$ and ${\bf 2}\xleftarrow[k_4]{}{\bf 1}$ respectively.

Obviously, large electrical energy $U$ and low external torque $\Gamma$ will result in high forward transition rates $k_1$ and $k_3$, but result in low backward transition rates $k_2$ and $k_4$, and consequently will lead to high rotation speed. Similar as in \cite{Lo2013}, rates $k_1^0, k_4^0$ depend on external and internal sodium concentrations [Na]$_{ex}$ and [Na]$_{in}$ respectively. For simplicity, this study assumed that $k_1^0=\hat k_1^0{\rm [Na]}_{ex}$, and $k_4^0=\hat k_4^0{\rm [Na]}_{in}$ \cite{Fisher2001, Kolomeisky2007}.
As mentioned previously, electrical energy $U$ can be obtained by $U=qV_m$, with membrane voltage $V_m$ satisfying the formulation in Eq. (\ref{eq1}).

Given the transition rates $k_1, k_2, k_3, k_4$ (see Fig. \ref{FIGschematic}({\bf b})), the mean rotation speed $V$ and dispersion $D$ of BFM can be easily obtained (see \cite{Derrida1983, Fisher19991, Zhang20092})
\begin{equation}\label{eq6}
V=\frac{(k_1k_3-k_2k_4)\varphi}{k_1+k_2+k_3+k_4},\quad
D=\frac{\frac12(k_1k_3+k_2k_4)\varphi^2-V^2}{k_1+k_2+k_3+k_4}.
\end{equation}
From the above formulation of rotation speed $V$, the {\it stall torque} $\Gamma_s$, i.e. the torque under which the rotation speed $V$ vanishes, can be obtained ad follows,
\begin{equation}\label{eq7}
\begin{aligned}
\Gamma_s=&\frac{U}{\varphi}+\frac{k_BT}{\varphi}\ln\left(\frac{k_1^0k_3^0}{k_2^0k_4^0}\right)
=\frac{NqV_m}{2\pi}+\frac{Nk_BT}{2\pi}\ln\left[\frac{(1.318+{\rm [Na]}_{ex})\hat k_1^0k_3^0}{9.013k_2^0\hat k_4^0}\right].
\end{aligned}
\end{equation}
Meanwhile, the randomness of BFM rotation can be obtained by $r=2D/V\varphi$.
By fitting to data about the torque-speed relationship presented in \cite{Lo2013}, the model parameters used in this study are listed in TABLE \ref{Tab1}.

\subsection*{Analysis for multiple-stator BFM}
Generally, one BFM may include multiple stators. Let $\lambda_k(\Gamma), \mu_k(\Gamma)$ be the binding and unbinding rates of stators to and from BFM rotor, provided that there are $k$ binding stators. One can easily get the the following equations
\begin{equation}\label{eq8}
\begin{aligned}
\frac{dp_0}{dt}=&\mu_1(\Gamma)p_1-n\lambda_0(\Gamma)p_0,\cr
\frac{dp_n}{dt}=&[(M-n+1)\lambda_{n-1}(\Gamma)p_{n-1}+
(n+1)\mu_{n+1}(\Gamma)p_{n+1}]-[(M-n)\lambda_n(\Gamma)+n\mu_n(\Gamma)]p_n,\ \ \textrm{for\ \ } 1\le n\le M-1,\cr
\frac{dp_M}{dt}=&\lambda_{M-1}(\Gamma)p_{M-1}-M\mu_M(\Gamma)p_M.
\end{aligned}
\end{equation}
Where $p_n$ is the probability that BFM rotor is bound by $n$ stators, $M$ is the total number of stators, and
\begin{equation}\label{eq9}
\lambda_n(\Gamma)=k_1(\Gamma/(n+1))+k_4(\Gamma/(n+1)),\quad
\mu_n(\Gamma)=k_2(\Gamma/n)+k_3(\Gamma/n).
\end{equation}
Here, binding/unbinding of one stator to/from BFM rotor means that this stator will/won't be engaged in BFM rotation. This study assumes that all binding stators share external torque $\Gamma$ equally (see Eq. (\ref{eq9})), which is similar as the one used for describing cargo motion in cells by multiple motor proteins \cite{Lipowsky2005, Lipowsky2008, Zhang2009}. The expressions given in Eq. (\ref{eq9}) based on the assumption that, one stator in state {\bf 2} means it is bound to the BFM rotor, and conversely one stator in state {\bf 1} means it is unbound from the BFM rotor.

At steady state, one can show that \cite{Derrida1983, Zhang20093}
\begin{equation}\label{eq10}
p_n=\left[\prod_{i=1}^n\left(\frac{u_{i-1}}{w_i}\right)\right]p_0,\quad \textrm{for\ } 1\le n\le M,
\end{equation}
where $u_n=(M-n)\lambda_n,\quad w_n=n\mu_n$, and $p_0$ can be obtained by the normalization condition,
\begin{equation}\label{eq11}
p_0=\frac{1}{1+\sum_{n=1}^M\left[\prod_{i=1}^n\left(\frac{u_{i-1}}{w_i}\right)\right]}.
\end{equation}
From probability $p_n$, the mean binding number of stator can be obtained by $\langle n\rangle=\sum_{k=1}^Mnp_n$, and the mean rotation speed of BFM can be obtained by $\overline{V}=\sum_{k=1}^Mp_nV_n$, where $V_n$ is the rotation speed if there are $n$ stators which are bound to the BFM rotor. The mean {\it stall torque} $\overline{\Gamma}_s$ is obtained by $\overline{V}(\overline{\Gamma}_s)=0$.

\section*{Acknowledgments}
This study was supported by the Natural Science Foundation of China (Grant No. 11271083), 
and the National Basic Research Program of China (National \lq\lq973" program, project No. 2011CBA00804).


\newpage

\section*{Tables}
\begin{table}[!ht]
\caption{The parameter values of the two-state model used in this study, which are obtained by fitting the rotation speed formulation listed in Eq. (\ref{eq6}) to the experimental data measured in \cite{Lo2013}. The total number of free parameter is 10, since $\epsilon_4, \epsilon_4'$ can be obtained by the normalization condition, $\sum_{i=1}^4\epsilon_i=\sum_{i=1}^4\epsilon_i'=1$.}
\begin{tabular}{cc|cc|cc}
  \hline
  parameter & value (s$^{-1}$) & parameter & value & parameter & value \\
  \hline\hline
  $\hat k_1^0$ & 1993.2 & $\epsilon_1$ & 0.16 & $\epsilon_1'$ & 0 \\
  $k_2^0$ & 43.8 & $\epsilon_2$ & 0.54 & $\epsilon_2'$ & 0.76 \\
  $k_3^0$ & 324.0 & $\epsilon_3$ & 0.3 & $\epsilon_3'$ & 0.08 \\
  $\hat k_4^0$ & 8258.5 & $\epsilon_4$ & 0 & $\epsilon_4'$ & 0.16 \\
  \hline
\end{tabular}
\label{Tab1}
 \end{table}

\newpage

\section*{Figure Legends}

\end{document}